\documentstyle[12pt]{article}
\begin{document}
\setlength{\unitlength}{1mm}
\title{{\hfill {\small Alberta-Thy 09-98} } \vspace*{2cm} \\
Black Hole Entropy in Induced Gravity: Reduction to 2D Quantum Field
Theory on the Horizon}
\author{\\
Valeri Frolov\thanks{e-mail: frolov@phys.ualberta.ca}
${}^{1}$ and Dmitri Fursaev\thanks{e-mail:
fursaev@thsun1.jinr.ru}
${}^{2}$ \date{}}
\maketitle
\noindent  {
$^{1}${ \em
Theoretical Physics Institute, Department of Physics, \ University of
Alberta, \\ Edmonton, Canada T6G 2J1}
\\ $^{2}${\em Joint Institute for
Nuclear Research,
Bogoliubov Laboratory of Theoretical Physics, \\
141 980 Dubna, Russia}
}
\bigskip

\begin{abstract}
It is argued that degrees of freedom responsible for
the Bekenstein-Hawking
entropy of a black hole in induced gravity are described by two
dimensional quantum field theory defined on the bifurcation surface of
the horizon.  This result is proved for a class of induced gravity
models with scalar, spinor and vector heavy constituents.
\end{abstract}

\bigskip\bigskip

\baselineskip=.6cm

\newpage

\section{Introduction}
\setcounter{equation}0

The statistical-mechanical origin of the Bekenstein-Hawking entropy
\cite{Hawk:75}, \cite{Beke:72} is one of the most intriguing problems
of the black hole physics. There exist several promising approaches to
this problem: string theory approach (see for a review Ref.
\cite{Peet:97}), calculations of the entropy of some 3D black holes
\cite{Carlip:95}, \cite{Strominger:97}, an explanation in the framework
of the loop quantum gravity \cite{ABCK:97}, a mechanism suggested in
Sakharov's induced gravity \cite{Sakharov} and others.
In the induced
gravity approach \cite{Jacobson}--\cite{FF:97} the Bekenstein-Hawking
entropy is related to the statistical-mechanical entropy of heavy
constituent fields which induce the Einstein theory in the low-energy
limit. Gravitons in the induced gravity are analogous to
phonon excitations in condensed matter systems \cite{Volovik:98}.

A special class of  induced gravity  models was investigated in Refs.
\cite{FFZ}, \cite{FF:97}. These models contain heavy spinors and
scalar constituents propagating in an external
gravitational field.  The dynamics of the gravitational field arises as
the result of quantum effects. The one-loop effective action for
quantum constituents gives the low  energy classical action for
the Einstein gravity. The constructed models of induced gravity are
free from the leading ultraviolet divergences. The induced Newton
constant $G$ is completely determined by the parameters of the
constituents, and it is finite only if non-minimally coupled scalar
fields are present. It was demonstrated  that  the  Bekenstein-Hawking
entropy $S^{BH}$ in the induced gravity can be written as
\begin{equation}\label{i1}
S^{BH}={{\cal A}\over 4G}=S^{SM}-Q\, .
\end{equation}
Here ${\cal A}$ is the surface area of the horizon, and
$S^{SM}=-\mbox{Tr}(\hat{\rho}\ln\hat{\rho})$ is the
statistical-mechanical (or entanglement) entropy of the thermally
excited (with thermal density matrix $\hat{\rho}$)  constituent fields
propagating near the horizon. The quantity
\begin{equation}\label{i1a}
Q=\sum_s\xi_s \int_{\Sigma}\langle \hat{\phi_s}^2\rangle
\end{equation}
is the sum of contributions of the nonminimally coupled scalar fields
$\hat{\phi_s}$. In this relation $\xi_s$ are parameters of nonminimal
coupling  and $\langle \hat{\phi_s}^2\rangle$ is the
quantum average of the  squares of the scalar operators on the
bifurcation surface $\Sigma$. In these particular models the origin of
$Q$ is related to the nonminmimal couplings of the scalar fields. It
was shown in \cite{FF:97}, $Q$ can be interpreted as a Noether charge
\cite{Wald:93}--\cite{JKM:94} and it determines
the difference between the energy of
the fields and their canonical Hamiltonian.

The subtraction in (\ref{i1}) is unavoidable for the following reasons.
The contribution of each (Bose and Fermi) constituent field  into
$S^{SM}$ is positive and divergent. Thus, the entropy $S^{SM}$ is
divergent, while black hole entropy $S^{BH}$ is finite. In formula
(\ref{i1}) the divergence of $S^{SM}$ is exactly compensated by the
divergence of the quantity $Q$.

There is a more profound reason why the Noether charge $Q$ appears in
(\ref{i1}). The Bekenstein-Hawking entropy $S^{BH}$ determines the
degeneracy of states of a black hole.  It was argued in \cite{FF:97}
that this degeneracy  can be calculated by  counting states  of
constituents with fixed total {\em energy}. On the other hand, the
entropy $S^{SM}$  is directly related to the distribution over the
levels of the {\em Hamiltonian} of constituent fields. The
additional term $Q$ is required to relate it to the distribution over
the energy levels.

\bigskip

In the present work we consider a wider class of induced gravity models
which besides scalar and spinor constituents contains also  massive
vector fields. For briefness we call such models {\em vector models}.
We demonstrate that the parameters of vector models can be chosen
so that to
exclude the leading ultraviolet divergences even if all scalar fields
are minimally coupled. The remarkable fact is that the relation
(\ref{i1}) is still valid. The Noether charge $Q$ in
equation (\ref{i1})  is related now to the ``natural'' coupling
of vector fields with the curvature. The universality of the form
of  Eq. (\ref{i1}) seams to be quite general property of the induced
gravity theories.

The important property of a vector model is that its only free
parameters are the masses of the fields, while the "nonminimal
couplings" are fixed by the form of the action of the vector fields. As
we will see, this property makes it possible a new interesting
interpretation of the Bekenstein-Hawking entropy in induced gravity in
terms of a two-dimensional quantum theory on $\Sigma$. Thus, the
induced gravity models provide a simple realization of the holographic
principle: the black hole entropy is encoded in  "surface" degrees of
freedom, i.e., in the degrees of freedom of the theory which propagate
very close to the black hole horizon. The holographic principle
was formulated in
\cite{Hooft:93}, \cite{Susskind} (see also recent paper
\cite{Hooft:98}) and at the present moment it is actively
discussed in the framework of the string theory
\cite{Maldacena:97}--\cite{SuWi:98}.

The paper is organized as follows. In Section 2 we describe
the models of induced gravity with vector fields.
Section 3 is devoted to the derivation of Eq. (\ref{i1}) for these
models.
Special attention here is paid to the calculation of
statistical-mechanical entropy of vector fields in the presence of
the Killing horizon
and to the properties of the Noether charge which is connected with
nonminimal vector couplings. These results
enable us to adopt the statistical-mechanical
explanation of the Bekenstein-Hawking entropy given in Ref. \cite{FF:97}
to more general class of induced gravity models.
In Section 4 we establish the relation between
the Bekenstein-Hawking entropy
and the effective action of a 2D free massive quantum fields ``living''
on the bifurcation surface $\Sigma$ of the horizons.
As we show this relation is satisfied for induced gravity
obtained from a theory with partly broken supersymmetry.
Concluding remarks and a brief discussion of the holographic
property of the black hole entropy in induced
gravity theories are presented in Section 5.
The relation between energy, Hamiltonian, and the Noether charge
for massive vector fields is derived in Appendix.

We use sign conventions of book \cite{MTW:73} and, thus,
we work with the signature $(-+++)$ for the Lorentzian metric.

\section{Induced gravity models with vector fields}
\setcounter{equation}0

The vector model\footnote{A similar model of induced gravity was discussed in
\cite{Zelnikov:97}.}
consists of $N_s$
minimally coupled
scalar fields $\phi_i$ with masses $m_{s,i}$, $N_d$ spinors $\psi_{j}$
with masses
$m_{d,j}$, and $N_v$ vector fields $V_k$ with masses
$m_{v,k}$. The classical actions of the fields are standard
\begin{equation}\label{1.1a}
I_{s}[\phi_i]=-\frac 12\int dV\left[(\nabla \phi_i)^2+
m_{s,i}^2\phi^2_i\right]~~~,
\end{equation}
\begin{equation}\label{1.2a}
I_{d}[\psi_j]= \int dV \bar{\psi}_{j}
(\gamma^\mu\nabla_\mu+m_{d,j})\psi_j~~~,
\end{equation}
\begin{equation}\label{1.3a}
I_{v}[V_k]=-\int dV\left[\frac 14 F_k^{\mu\nu}F_{k\mu\nu}+
\frac 12 m_{v,k}^2V_k^\mu V_{k\mu}\right]~~~,
\end{equation}
where $dV=\sqrt{-g}d^4x$ is the volume element of 4D space-time ${\cal
M}$ and $F_{k\mu\nu}=\nabla_\mu V_{k\nu}-\nabla_\nu V_{k\mu}$. The
corresponding quantum effective action of the model is
\begin{equation}\label{1.4a}
\Gamma=\sum_{i=1}^{N_s}\Gamma_s(m_{s,i})+\sum_{j=1}^{N_d} \Gamma(m_{d,j})+
\sum_{k=1}^{N_v} \Gamma(m_{v,k})~~~.
\end{equation}
$\Gamma$ is a functional of the metric $g_{\mu\nu}$ of the background
space-time.
The scalar and spinor actions follow immediately from
Eqs. (\ref{1.1a}) and (\ref{1.2a}),
\begin{equation}\label{1.5a}
\Gamma_s(m_{s,i})=\frac 12 \log\det(-\nabla^2+m_{s,i}^2)~~~,
\end{equation}
\begin{equation}\label{1.6a}
\Gamma_d(m_{d,j})=-\log\det(\gamma^\mu\nabla_\mu+m_{d,j})~~~.
\end{equation}
As a result of equation of motion, a massive vector field $V_\mu$ obeys
the condition $\nabla^\mu V_\mu=0$, which leaves only
three independent components. Under quantization this condition
can be realized as a constraint so that the effective action
for vector fields takes the form
\begin{equation}\label{1.7a}
\Gamma_v(m_{v,k})=\tilde{\Gamma}_v(m_{v,k})- \Gamma_s(m_{v,k})~~~,
\end{equation}
\begin{equation}\label{1.8a}
\tilde{\Gamma}_v(m_{v,k})=
\frac 12 \log\det(-\nabla^2\delta^\mu_\nu+R^\mu_\nu+
m_{v,k}^2\delta^\mu_\nu)~~~,
\end{equation}
where $R^\mu_\nu$ is the Ricci tensor. The functional
$\tilde{\Gamma}_v(m_{v,k})$
represents the effective action for a massive vector field
which we will denote as $A_{k,\mu}$. The classical action
for $A_{k,\mu}$ which results in (\ref{1.8a}) is
\begin{equation}\label{1.9a}
\tilde{I}_{v}[A_k]=-\frac 12\int dV\left[\nabla^{\mu}A_k^\nu
\nabla_\mu A_{k\nu}+R_{\mu\nu}A_k^\mu A_k^\nu+
m_{v,k}^2A_k^\mu A_{k\mu}\right]~~~.
\end{equation}
The field $A_k^\mu$ obeys no constraints and carries an
extra degree of freedom.
The contribution of this unphysical degree of freedom
in (\ref{1.8a}) is compensated by subtracting the
action $\Gamma_s(m_{v,k})$ of a scalar field with the mass $m_{v,k}$,
see Eq. (\ref{1.7a}).

\bigskip

In general, the effective action (\ref{1.4a}) is ultraviolet
divergent quantity. Let us discuss now the constraints which have
to be imposed on the masses of the constituents to eliminate
the leading divergences in $\Gamma$.
The divergences related to
each particular field follow from the Schwinger-DeWitt
representation
\begin{equation}\label{1.10a}
\Gamma_i=-{\eta_i \over 2}\int^{\infty}_{\delta}{ds \over s} e^{-m_i^2s}
\mbox{Tr} e^{-s\Delta_i}~~~,
\end{equation}
where $\eta_i=+1$ for Bose fields and $-1$ for Fermi fields, and
$\delta$ is an ultraviolet cutoff. The divergences come from the
lower integration limit where one can use the asymptotic expansion
of the trace of the heat kernel operator of $\Delta_i$
\begin{equation}\label{1.11a}
\mbox{Tr} e^{-s\Delta_i}\simeq {1 \over (4\pi s)^2}\int dV
(a_{i,0}+sa_{i,1}+...)~~~.
\end{equation}
For the fields under consideration we have
\begin{equation}\label{1.12a}
\Delta_s=-\nabla^\mu\nabla_\mu~~,~~a_{s,0}=1~~,~~a_{s,1}=\frac 16 R~~~,
\end{equation}
\begin{equation}\label{1.13a}
\Delta_d=-(\gamma^\mu\nabla_\mu)^2~~,~~a_{d,0}=4~~,~~a_{d,1}=-\frac 13 R~~~,
\end{equation}
\begin{equation}\label{1.14a}
(\Delta_v)^\mu_\nu=-\nabla^\rho\nabla_\rho\delta^\mu_\nu
+R^\mu_\nu~~,~~a_{v,0}=4~~,~~a_{v,1}=-\frac 13 R~~.
\end{equation}
As in the case of the model considered in Ref. \cite{FFZ}, we require
vanishing of the cosmological
constant and cancelation of the divergences of the
induced Newton constant.
These conditions can be written down with the help of the
following two functions
\begin{equation}\label{1.5}
p(z)=\sum_{i=1}^{N_s} m_{s,i}^{2z}-4\sum_{j=1}^{N_d} m_{d,j}^{2z}+
3\sum_{k=1}^{N_v} m_{v,k}^{2z}~~~,
~~~
q(z)=\sum_{i=1}^{N_s} m_{s,i}^{2z}+2\sum_{j=1}^{N_d} m_{d,j}^{2z}-3
\sum_{k=1}^{N_v} m_{v,k}^{2z}~~~.
\end{equation}
As can be shown by using Eqs. (\ref{1.4a}), (\ref{1.10a})--(\ref{1.14a}),
the induced cosmological constant vanishes when
\begin{equation}\label{1.6}
p(0)=p(1)=p(2)=p'(2)=0~~~.
\end{equation}
The induced Newton constant $G$ is finite
if
\begin{equation}\label{1.7}
q(0)=q(1)=0~~~.
\end{equation}
The constraints
result in simple relations
\begin{equation}\label{1.6aa}
N_s=N_d=N_v~~~,~~~\sum_{i=1}^{N_s} m_{s,i}^2=\sum_{j=1}^{N_d} m_{d,j}^2=
\sum_{k=1}^{N_v} m_{v,k}^2~~~.
\end{equation}
They show that one cannot construct the theory
with finite cosmological and Newton constants
from vector and spinor fields only.

\bigskip

The low-energy limit of the theory corresponds to the regime when the
curvature radius $L$ of the spacetime ${\cal M}$ is much greater than
the Planck length $m_{Pl}^{-1}$. In this limit the effective action
$\Gamma$ of the theory can be expanded in the curvature. The terms in
this series are local and the leading terms  can be calculated
explicitly. In the linear in curvature approximation $\Gamma$ coincides
with the Einstein action \footnote{To induce the correct boundary term
in (\ref{1.15a}) one has to add to $\Gamma$ an integral of
averages of field operators on the spatial boundary ${\partial {\cal
M}}$, see \cite{BaSo:96}. These terms are not relevant for our
analysis. Let us emphasize that we are interested in the
statistical-mechanical computation of the black hole entropy for which
only the region near the horizon is important.}
\begin{equation}\label{1.15a}
\Gamma[g]\simeq{1 \over 16\pi G}\left(\int_{\cal M}dV R
+2\int_{\partial{\cal M}}dv K\right)~~~.
\end{equation}
Here $dv$ is the volume element of $\partial {\cal M}$. The Newton
constant is determined by the following expression
\begin{equation}\label{1.8}
{1 \over G}={1 \over 12\pi}q'(1)=
{1 \over 12\pi}\sum_{i=1}^{N}\left( m_{s,i}^2\ln m_{s,i}^2+
2m_{d,i}^2\ln m_{d,i}^2
-3m_{v,i}^2\ln m_{v,i}^2
\right)~~~.
\end{equation}
Here, according to (\ref{1.6aa}), we put $N=N_s=N_d=N_v$. From this
expression it is easy to conclude that at least some of the
constituents must be heavy and have mass comparable with the Planck
mass $m_{Pl}$. For simplicity in what follows we assume that all the
constituents are heavy.

\bigskip

Let us analyze models where conditions (\ref{1.5}) and (\ref{1.6})
are satisfied. Equations (\ref{1.6aa}) are trivially satisfied when all
fields are in supersymmetric multiplets. However  in such
supersymmetric models $p(z)=q(z)\equiv 0$ (because masses of the fields
in the same supermultiplet coincide) and the induced gravitational
constant vanishes. A nontrivial induced  gravity theory can be obtained
if the supersymmetry is partly broken by splitting the masses of the
fields in the supermultiplets.

Let us demonstrate this by an example. Consider the model with $N$
massive supermultiplets. Each multiplet consists of one scalar, one
Dirac spinor and one vector field, so that the numbers of Bose and
Fermi degrees of freedom coincide\footnote{Supersymmetric models with
free massive scalar, spinor and vector fields are discussed, for
instance, in Ref. \cite{LoWo:81}}. We suggest that masses of vector and
spinor fields are equal, $m_{v,i}=m_{d,i}\equiv m_i$ (here $i$ is the
number of the multiplet). The masses of the scalar partners are assumed
to be $m_{s,i}=(1+x_i)m_i$, where $x_i$ is a dimensionless coefficient.
The case when $|x_i|\ll 1$ corresponds to slightly broken
supersymmetry. For this case
\begin{equation}\label{1.16a}
p(z)=q(z)=\sum_{i=1}^N m_i^{2z}\left[(1+x_i)^{2z}-1\right]\simeq
2z\sum_{i=1}^N x_i m_i^{2z}~~~.
\end{equation}
Now equations (\ref{1.6}), (\ref{1.7}), and (\ref{1.8}) take the simple
form
\begin{equation}\label{1.17a}
\sum_{i=1}^N x_i m_i^2=0~~~,~~~\sum_{i=1}^N x_i m_i^4=0~~~,
\end{equation}
\begin{equation}\label{1.18a}
\sum_{i=1}^N x_i m_i^4\ln m_i^2=0~~~,~~~{1 \over G}\simeq{1 \over 6\pi}
\sum_{i=1}^N x_i m_i^2\ln m_i^2~~~.
\end{equation}
This is a system of linear equations for $x_i$
which for $N\ge 4$ has nontrivial
solutions.

The induced gravity constraints provide cancelation of the leading
ultraviolet divergences. However, some logarithmical divergences are
still present on general backgrounds. On the Schwarzschild background
the logarithmic divergences are pure topological and can be neglected.
That is why in what follows we restrict the analysis to black holes of
this type\footnote{ At least some of the logarithmic divergences can
be eliminated in more complicated models, for instance in models which
contain both vector  and non-minimally coupled scalar fields. These
models allow one to generalize the analysis of the black hole entropy
problem in induced gravity to charged black holes.}.

\section{Statistical calculation of the
black hole entropy}
\setcounter{equation}0

Let us now calculate the statistical-mechanical entropy $S^{SM}$ in
the vector models of induced gravity and compare it with the
Bekenstein-Hawking entropy of a black hole. As a result of this
comparison, we prove the validity of  equation (\ref{i1}) for these
models.

The canonical ensemble of constituent fields on a static,
asymptotically flat background can be described by standard methods.
The statistical-mechanical entropy of the fields
is determined from the free energy
\begin{equation}\label{2.1a}
F(\beta)=-\beta^{-1}\ln \mbox{Tr} \exp(-\beta :\!\hat{H}\!:)
=\eta \beta^{-1}\int_0^{\infty}d\omega {dn \over d\omega}
\ln\left(1-\eta e^{-\beta\omega}\right)~~~.
\end{equation}
Here $\beta$ is the inverse temperature measured at infinity and
$:\!\hat{H}\!:$ is the Hamiltonian of the system which is defined as
the generator of canonical transformations along the Killing time. The
factor $\eta= 1$ for bosons and $\eta= -1$ for fermions,  $\omega$ are
the frequencies of single-particle excitations, $dn/ d\omega$ is the
density of levels $\omega$.

When the background space-time is the exterior region of a black hole
the single particle-spectra have a number of specific properties
because of the presence of the Killing horizons \cite{FF:98}.  In
particular, the density of states $dn/ d\omega$ infinitely grows near
the horizon. Although this divergence has an infrared origin,
regularizations of the ultraviolet type  can be applied to make $dn
/d\omega$ finite. For scalar and spinor fields on general static
backgrounds the divergences of $dn /d\omega$ were computed in
\cite{Fursaev:97}. In the Pauli-Villars regularization the leading
divergences for scalar and Dirac spinor fields of the mass $m$ are
\begin{equation}\label{2.2a}
{dn_s(m) \over d\omega}= {b(m) \over 8\pi^2\kappa}\,  {\cal A}~~~,
~~~
{dn_d(m) \over d\omega}= {b(m) \over 2\pi^2\kappa} \,  {\cal A}~~~,
\end{equation}
\begin{equation}\label{2.3a}
b(m)=c\mu^2-m^2\ln{\mu^2 \over m^2}~~~.
\end{equation}
Here $\kappa=(4M)^{-1}$ is the surface gravity of the black hole,
$\mu$ is the Pauli-Villars cutoff, and $c=\ln{729 \over 256} >0$.

Modes  propagating in the vicinity of the horizon give the main
contribution to the densities of levels. That is why the quantity $dn
/d\omega$ scales as the surface area ${\cal A}$ of the horizon. This
also means that to get (\ref{2.2a}) it is sufficient to restrict
oneself by the Rindler approximation of the black hole metric
\begin{equation}\label{1.9aa}
ds^2=-\kappa^2\rho^2dt^2+d\rho^2+dz_1^2+dz_2^2~~~.
\end{equation}
Here $\rho>0$, and $t$ is the Rindler  time coordinate. In this
approximation  the densities of levels for high spin fields can be
computed by using expression (\ref{2.2a}) for scalars and spinors.

Let us consider a massive vector field in the Minkowski spacetime.
We denote by $X^{m}$ ($m=0,\ldots, 4$) the Cartesian coordinates in
this space and by $V_m=(V_0,V_a)$, $a=1,2,3$,
the components of the vector field with respect to the Cartesian
frame. Then the equations of motion which extremize vector
field action (\ref{1.3a}) are simply a set of Klein-Gordon equations
for four ``scalars'' $V_m$ plus the additional constraint $\partial_m
V^m=0$. The constraint serves to express the time-component $V_0$ in
terms of other components $V_a$. The contribution of this component to
the energy is negative and $V_0$ cannot be considered as an independent
physical degree of freedom. The density of levels of the vector field
$dn_v /d\omega$ multiplied by $d\omega$ is the number of independent
solutions $V_m(t,\rho,z)=e^{-i\omega t}V_m(\rho,z)$ of the field
equations with the frequencies in the interval $(\omega,
\omega+d\omega)$. The solutions are determined by three independent
functions $V_\alpha$. Therefore, in the Rindler approximation $dn_v /
d\omega$  is greater by the factor 3 than the density of levels of a
scalar field of the same mass $m$. From (\ref{2.2a}) we find
\begin{equation}\label{2.4a}
{dn_v(m) \over d\omega}=3{dn_s(m) \over d\omega}=
{3b(m) \over 8\pi^2\kappa} {\cal A}~~~.
\end{equation}
The curvature corrections may change this relation
but they are not important for further analysis.

The statistical-mechanical entropy
\begin{equation}\label{2.5ab}
S=\beta^2 {\partial F \over \partial \beta}~~,~~
\end{equation}
of scalar, spinor and vector fields follows from Eqs. (\ref{2.1a}),
(\ref{2.2a}) and (\ref{2.4a})
\begin{equation}\label{2.5a}
S_s(m_{s,i})={b(m_{s,i}) \over 48\pi}{\cal A}~~,~~
S_d(m_{d,i})={2b(m_{d,i}) \over 48\pi}{\cal A}~~,~~
S_v(m_{v,i})={3b(m_{v,i}) \over 48\pi}{\cal A}~~.
\end{equation}
Expressions (\ref{2.5a}) are obtained from formula (\ref{2.5ab})
at the Hawking temperature, i.e., at $\beta=2\pi/\kappa=8\pi M$.
The statistical-mechanical entropy of the constituents in
the induced gravity model is
\begin{equation}\label{2.6a}
S^{SM}=\sum_{i=1}^N\left[S_s(m_{s,i})+S_d(m_{d,i})+S_v(m_{v,i})\right]~~~.
\end{equation}
By substituting (\ref{2.5a}) into (\ref{2.6a})
and taking into account (\ref{2.3a}), (\ref{1.6aa})
we get
$$
S^{SM}={1 \over 48\pi}\sum_{i=1}^N\left[m_{s,i}^2\ln m_{s,i}^2+
2m_{d,i}^2\ln m_{d,i}^2+3m_{v,i}^2\ln m_{v,i}^2\right]{\cal A}
$$
\begin{equation}\label{2.7a}
+{1 \over 8\pi}\left[c N\mu^2-\ln \mu^2
\sum_{i=1}^N m_{v,i}^2\right]{\cal A}~~~.
\end{equation}

Let us now calculate the Noether charge $Q$ for our model.
It is instructive to
discuss first the entropy of a black hole in a classical theory. According
to Wald et al. \cite{Wald:93}--\cite{JKM:94}, the black hole entropy
can be interpreted as a Noether charge and obtained from the
Lagrangian $L$ of the theory. For theories which do not include
the derivatives of the metric higher than second order the entropy
can be written in the form
\begin{equation}\label{2.8a}
S=-8\pi\int_{\Sigma} t_\mu n_\nu t_\lambda n_\rho
{\partial L \over
\partial R_{\mu\nu\lambda\rho}}d\sigma~~~,
\end{equation}
where $R_{\mu\nu\lambda\rho}$ is the Riemann tensor. The integration
in (\ref{2.8a})
goes over the bifurcation surface $\Sigma$ of the horizon, $d\sigma$
is the volume element of $\Sigma$ ($\int_{\Sigma} d\sigma={\cal A}$).
Vectors $t_\mu$ and $n_\mu$ are two mutually orthogonal
vectors normal to $\Sigma$ such that $t^2=-1$ and $n^2=1$.

For the Einstein theory equation (\ref{2.8a}) reproduces the
Bekenstein-Hawking formula for the black hole entropy. The important
consequence of (\ref{2.8a}) is that coupling of the
matter fields with the curvature  gives a nonzero contribution
$\Delta S$ to the Bekenstein-Hawking entropy. In quantum theory $\Delta
S$ becomes an average of the corresponding field operator on $\Sigma$.

Let us now consider the vector model of induced gravity. According to
(\ref{1.4a}) the effective action of the theory can be written as a
path integral
\begin{equation}\label{2.9a}
\exp{i\Gamma[g]}=\int [D\Phi]\exp (iI[g,\Phi])~~~,
\end{equation}
\begin{equation}\label{2.10a}
I[g,\Phi]\equiv\sum_{i=1}^N\left(I_s[\phi_i]+I_d[\psi_i]+\tilde{I}_v[A_i]
+I_s[\varphi_i]\right)~~~,
\end{equation}
where $\Phi=\{\phi_i,\psi_i, A_i,\varphi_i\}$. The
functionals $I_s$, $I_d$, and $\tilde{I}_v$ are defined by
Eqs. (\ref{1.1a}), (\ref{1.2a}), and (\ref{1.9a}), respectively.
The origin of the scalar fields $\varphi_i$ in (\ref{2.9a})
is related to the quantization of the
massive vector fields $A_i$. It is assumed that $\varphi_i$
obey the ``wrong'' (Fermi) statistics in order to reproduce Eq. (\ref{1.7a}).
As follows from
(\ref{1.9a}), the total ``classical action'' $I[g,\Phi]$ includes
the nonminimal couplings of the vector fields $A_i$. By using
formula (\ref{2.8a}) in the theory with the action $\tilde{I}_v[A]$
one obtains the nonzero term
\begin{equation}\label{2.11a}
\Delta S=-8\pi\int_{\Sigma} t_\mu n_\nu t_\lambda n_\rho
{\partial \tilde{L}_v \over
\partial R_{\mu\nu\lambda\rho}}d\sigma=\pi
\int_{\Sigma}(t^\mu t^\nu-n^\mu n^\nu) A_\mu A_\nu d\sigma~~~,
\end{equation}
where we put $\tilde{I}_v[A]=\int \tilde{L}_v dV$.
In the induced gravity such terms result in a correction
to the entropy of a black hole. In the first order in the Planck constant
this correction simply is
\begin{equation}\label{2.12a}
\Delta S=\pi
\int_{\Sigma}(t^\mu t^\nu-n^\mu n^\nu) \sum_{i=1}^N
\langle \hat{A}_{i\mu} \hat{A}_{i\nu} \rangle d\sigma\equiv
-Q~~~.
\end{equation}
Here the average $\langle \hat{A}_{i\mu} \hat{A}_{i\nu} \rangle$ is
understood as a regularized quantity. The quantity $Q$ has a meaning of
Wald's Noether charge associated with nonminimal interaction terms of
the vector field. The sign minus in the r.h.s. of (\ref{2.12a}) is chosen
so that $Q$ be positive.

By using the Pauli-Villars regularization one finds that in the Rindler
approximation
\begin{equation}\label{2.13a}
\langle \hat{A}_{i\mu} \hat{A}_{i\nu} \rangle=\eta_{\mu\nu}
{b(m_{v,i}) \over 16\pi^2}~~~,
\end{equation}
where $\eta_{\mu\nu}$ is the Minkowski metric and function $b(m_{v,i})$
is defined by (\ref{2.3a}). Equation (\ref{2.13a}) gives
\begin{equation}\label{2.14a}
Q={1 \over 8\pi}\sum_{i=1}^N b(m_i) {\cal A}={1 \over 8\pi}
\left(cN\mu^2-\ln \mu^2\sum_{i=1}^N m_{v,i}^2+
\sum_{i=1}^N m_{v,i}^2\ln m_{v,i}^2
\right){\cal A}~~~.
\end{equation}

This result allows one to show that the total Bekenstein-Hawking
entropy $S^{BH}$ in induced gravity is the difference of
statistical-mechanical entropy $S^{SM}$, see Eq. (\ref{2.7a}), and the
Noether charge $Q$. As can be easily seen, the divergences of $S^{SM}$
are exactly canceled by the divergences of the charge $Q$, so that one
gets the finite expression
\begin{equation}\label{2.15a}
S^{SM}-Q={1 \over 48\pi}\sum_{i=1}^N\left[m_{s,i}^2\ln m_{s,i}^2+
2m_{d,i}^2\ln m_{d,i}^2-3m_{v,i}^2\ln m_{v,i}^2\right]{\cal A}=
{ {\cal A}\over 4G}=S^{BH}~~~.
\end{equation}
This expression coincides exactly with the Bekenstein-Hawking entropy
in induced gravity where the induced Newton constant is determined by
formula (\ref{1.8}).

As was argued in Ref. \cite{FF:97}, the statistical-mechanical reason
why the Noether charge appears in (\ref{2.15a}) is related to the fact
that the canonical Hamiltonian $H$ and the energy $E$ of the system are
different. $H$ defines free energy (\ref{2.1a}) and entropy $S^{SM}$
while the energy $E$ is connected with the spectrum of the mass of the
black hole.
In Appendix we show that for the vector model
\begin{equation}\label{2.21a}
H-E={\kappa \over 2\pi} Q~~~.
\end{equation}
This is the same relation which was found in \cite{FF:97} for the
induced gravity model with non-minimally coupled scalar fields.
Relation (\ref{2.21a}) can be used to provide the
statistical-mechanical interpretation of the subtraction of the charge
$Q$ in the black hole entropy formula (\ref{2.15a}). This
interpretation repeats the one already given in \cite{FF:97}: the
subtraction of $Q$ is needed in order to pass from the distribution over
the energy in canonical ensemble of constituent fields to the
distribution over the black hole mass in the black hole canonical
ensemble which determines $S^{BH}$.

\section{Black hole entropy and 2D quantum theory on $\Sigma$}
\setcounter{equation}0

Our aim now is to relate the Bekenstein-Hawking entropy $S^{BH}$,
Eq. (\ref{2.15a}), to a 2D quantum theory of free massive fields
``living'' on the bifurcation surface $\Sigma$ of the horizon.
To this aim it is instructive to represent expression (\ref{2.15a})
in another equivalent form. First, let us note that in the
Rindler approximation  the regularized correlators of the
scalar, spinor and vector fields of the mass $m$ have the simple form
\begin{equation}\label{1.9}
\langle \hat{\phi}^2 \rangle={b(m) \over 16\pi^2}~~,~~
\langle \hat{\bar{\psi}}\hat{\psi}\rangle=4m\langle \hat{\phi}^2 \rangle
~~,~~
\langle \hat{V}_\mu\hat{V}^\mu \rangle=
3\langle \hat{\phi}^2 \rangle~~~.
\end{equation}
In the Pauli-Villars regularization the function $b$ is defined by
(\ref{2.3a}). From Eqs. (\ref{2.15a}) and (\ref{1.9})
we easily find that
\begin{equation}\label{1.14}
S^{BH}={\pi\over 6}\sum_{i=1}^{N}\int_{\Sigma}d\sigma \left[
2\langle \hat{\phi}_{i}^2 \rangle
+{1 \over m_{d,i}}
\langle \hat{\bar{\psi}}_{i}\hat{\psi}_{i}\rangle
-2\langle \hat{V}^2_{i} \rangle
\right]~~~.
\end{equation}
One can check that
the divergences in correlators in (\ref{1.14})
are canceled because of induced gravity
constraint $q(1)=0$, see Eqs. (\ref{1.5}) and (\ref{1.7}).
Since the surface $\Sigma$ of bifurcation of horizons is a set of fixed
points of the Killing vector, only zero-frequency (``soft'') modes
contribute to the correlators on $\Sigma$ (for a detailed discussion of
this point, see \cite{FF:97}).

As was shown in \cite{FF:97}, the correlator of scalar
fields taken on the bifurcation surface of the Killing horizons
behaves effectively as a two-dimensional operator.
Namely, if $z$ and $z'$ are the coordinates
of the points $x$ and $x'$ on $\Sigma$, see (\ref{1.9aa}), then
\begin{equation}\label{4.1}
\langle\hat{\phi}(x(z))\hat{\phi}(x(z'))
\rangle=-
{1 \over 4\pi}
\langle z|\ln O_{\Sigma}|z'\rangle~~~,
\end{equation}
\begin{equation}\label{4.3}
O_{\Sigma}=-\nabla^2_{\Sigma}+m^2~~~,
\end{equation}
where $-\nabla^2_{\Sigma}$ is the Laplacian on $\Sigma$. The left and
right parts of (\ref{4.1}) should be calculated in the same
regularization. It should be emphasized that (\ref{4.1}) is exact
relation for the Rindler space\footnote{It can be generalized to curved
backgrounds with a Killing horizon. In general case the operator
$O_\Sigma$ for very massive fields can be found by comparing
Schwinger-DeWitt asymptotics of four and two dimensional operators, see
for the details \cite{FF:97}. The key property which allows two
dimensional interpretation of 4D correlators on $\Sigma$ is that
$\Sigma$ is a geodesic surface. That is, any 4D geodesic which begins
and ends on $\Sigma$ coincides with 2D geodesic on $\Sigma$.}. For r.h.s.
of (\ref{4.1}) we find that
\begin{equation}\label{4.7}
-{1 \over 4\pi} \int_{\Sigma}\langle z|\ln O_{\Sigma}|z \rangle ~
d\sigma=-{1 \over 4\pi} \ln \det(
-\nabla^2_{\Sigma}+m^2)
\equiv -{1 \over 2\pi} W_s(m)~~~.
\end{equation}
The functional $W_s(m)$ has the meaning of the effective action of a 2D
quantum field $\chi$ given on $\Sigma$. It can be expressed in terms of
the Euclidean path integral as
\begin{equation}\label{a4.10}
e^{-W_s(m)}=\int D[\chi]
\exp\left[-\frac 12\int_{\Sigma}((\nabla_\Sigma\chi)^2+m^2\chi^2)
d\sigma\right]
~~~,
\end{equation}
where $D[\chi]$ is a covariant measure. In Ref. \cite{FF:97} a
two-dimensional auxiliary field $\chi$ ``living'' on the bifurcation
surface $\Sigma$ was called a {\it flucton} field to distinguish it from
the 4D fields in the black hole exterior. From (\ref{4.1}) and
(\ref{4.7}) one obtains
\begin{equation}\label{a4.9a}
\int_{\Sigma}d\sigma \langle \hat{\phi}^2(x(z))\rangle
=-{1 \over 2\pi}W_s(m)~~~.
\end{equation}
It follows from (\ref{1.14}) and (\ref{a4.9a}) that the contribution
of scalar fields to the black hole entropy $S^{BH}$ can be interpreted in
terms of 2D quantum theory of fluctons on $\Sigma$.

\bigskip

We now find an analogous representation for the
contribution to $S^{BH}$ from spinor and vector
fields. The correlators of these fields in the coinciding points
are tensors in the certain representations of the Lorentz group.
Different parts of these tensors  have different two-dimensional
interpretation.
We begin with the correlator (\ref{4.11}) of vector fields
restricted on $\Sigma$
\begin{equation}\label{4.11}
\langle\hat{V}_\mu(x(z))\hat{V}_\nu(x(z'))
\rangle=\langle\hat{A}_\mu(x(z))\hat{A}_\nu(x(z'))\rangle+
m^{-2}\nabla_\mu\nabla'_\nu\langle\hat{\varphi}(x(z))\hat{\varphi}(x(z'))
\rangle~~~,
\end{equation}
where $\hat{\varphi}$ is a scalar field of the same mass as
$\hat{V}_\mu$.
Let us consider the components of tensor quantities in
the Minkowski coordinates $X^m$. With respect to the coordinate
transformations on $\Sigma$, components of a tensor with indices $0$
and $3$ behave as scalars while components with indices $1$ and $2$
transform as vectors on $\Sigma$. By using arguments of Ref.
\cite{FF:97} one can express the correlator
$\langle\hat{A}_\mu\hat{A}_\nu\rangle$ with $\mu,\nu=1,2$ in terms of
the 2D vector Laplacian on $\Sigma$. Analogously, components of the
correlator with $\mu,\nu=0,3$ can be represented in terms of the 2D scalar
Laplacian. Thus, we find that
\begin{equation}\label{4.12}
\int_{\Sigma}d\sigma\langle
\hat{A}^1(x(z))\hat{A}_1(x(z))+\hat{A}^2(x(z))\hat{A}_2(x(z))
\rangle=-{1 \over 2\pi}\tilde{W}_v(m)~~~,
\end{equation}
\begin{equation}\label{4.13}
\int_{\Sigma}d\sigma\langle
\hat{A}^0(x(z))\hat{A}_0(x(z))+\hat{A}^3(x(z))\hat{A}_3(x(z))
\rangle=-2{1 \over 2\pi}W_s(m)~~~.
\end{equation}
$W_s$ is the effective action of a scalar field on $\Sigma$ and
$\tilde{W}_v(m)$ is the effective action of a vector field on $\Sigma$
with the same mass $m$ as that of 4D field
\begin{equation}\label{4.12ab}
\tilde{W}_v(m)=\frac 12 \log\det((-\nabla_{\Sigma}^2+\frac 12 R_\Sigma
+m^2)\delta^A_B)~~~.
\end{equation}
Here $A,B=1,2$ and $R_\Sigma$ is the curvature of $\Sigma$
which can be neglected in the Rindler approximation.
It follows from (\ref{4.11})--(\ref{4.13})
that
$$
\int_{\Sigma}d\sigma\langle \hat{V}^\mu(x(z))\hat{V}_\mu(x(z))
\rangle
$$
\begin{equation}\label{4.13a}
=\int_{\Sigma}d\sigma\langle \hat{A}^\mu(x(z))\hat{A}_\mu(x(z))
-\hat{\varphi}(x(z))\hat{\varphi}(x(z))\rangle
=-{1 \over 2\pi}(W_v(m)+2W_s(m))~~~,
\end{equation}
\begin{equation}\label{4.13ab}
W_v(m)=\tilde{W}_v(m)-W_s(m)~~~.
\end{equation}
The functional $W_v(m)$ corresponds
to the quantization of the massive 2D vector field described by
the classical action analogous to 4D action (\ref{1.3a}).

Similar relations can be obtained for spinor fields. One 4D
Dirac spinor corresponds to two 2D Dirac spinors on $\Sigma$. One
easily finds that
\begin{equation}\label{4.14}
\int_{\Sigma}d\sigma\langle \hat{\bar{\psi}}(x(z))\hat{\psi}(x(z))
\rangle={m \over \pi}W_d(m)~~~,
\end{equation}
where  $W_d(m)$ is the effective action of 2D spinors
on $\Sigma$ with mass $m$.

\bigskip

By using equations (\ref{a4.9a}), (\ref{4.13a}) and (\ref{4.14})
in expression (\ref{1.14}) for the Bekenstein-Hawking entropy in
induced gravity we find
\begin{equation}\label{2.10}
S^{BH}=\frac 16\sum_{i=1}^{N}\left[-
W_{s}(m_{s,i})+W_{d}(m_{d,i})+W_{v}(m_{v,i})+2W_{s}(m_{v,i})
\right]~~~.
\end{equation}
This form of the entropy looks similar to the effective action of a two
dimensional quantum field model on the surface $\Sigma$. To make this
similarity more evident let us consider the concrete induced gravity
model with partially broken supersymmetry which was discussed in
Section 2. In this model the masses of vector and spinor fields
coincide, $m_{v,i}=m_{d,i}=m_i$. As a result, $W_{v}(m)=W_{s}(m)=-\frac
12 W_d(m)$ and Eq. (\ref{2.10})  takes the form
\begin{equation}\label{2.10aa}
S^{BH}=-{1 \over 12} \sum_{i=1}^{N}
\left[2W_s(m_{s,i})+W_{d}(m_{i})
\right]\equiv -{1 \over 12}\Gamma^{(2)}~~~.
\end{equation}
The quantity $\Gamma^{(2)}$ is the effective action of a 2D
model which consists of $N$ spinor fields with masses $m_{i}$
and $2N$ scalar fields with
masses $m_{s,i}$.

In fact, we have 2D induced gravity on $\Sigma$.
The condition that the 4D curvature is small compared to the masses
of the fields guarantees that the two-dimensional curvature of
$\Sigma$ is small as well. So the
2D effective action $\Gamma^{(2)}$ can be computed as an expansion
in curvature. The leading term in this expansion is
the cosmological constant term
\begin{equation}\label{2.3}
\Gamma^{(2)}[\gamma]\simeq\int_{\Sigma} \sqrt{\gamma} \, \, d^2x\, \,
\lambda~~~.
\end{equation}
Here $\lambda$ is the ``induced'' 2D cosmological constant which
is expressed in terms of induced 4D Newton constant (\ref{1.8})
as $\lambda=-3/G$.
The constraints which provide the ultraviolet finiteness
of 4D Newton constant, see Eqs. (\ref{1.7}), automatically
guarantee the finiteness of 2D cosmological constant.

The 2D model described by functional $\Gamma^{(2)}$
can be obtained from
the supersymmetric model with $N$ multiplets
consisting of a spinor and 2 scalar fields. The split of the masses of
spinor and scalar fields breaks the supersymmetry and yields nonvanishing
2D cosmological constant.

Of course, the suggested connection between 4D and 2D theories is not
unique, and one may expect that in general the coefficient in r.h.s. of
(\ref{2.10aa}) can be another rational number. Let us emphasize that
the considered models of induced gravity  are phenomenological and
admit a large arbitrariness in the choice of masses of the constituent
fields. One may hope that if the induced gravity is obtained from an
underlying fundamental theory the masses of the fields will be fixed by
some principle which will determine the coefficient in (\ref{2.10aa}).

A remark is also in order about two-dimensional interpretation
of the Noether charge $Q$. By taking into account Eq. (\ref{2.14a}) it
is easy to show that in the Rindler approximation
\begin{equation}\label{xx}
Q=-\sum_{i}W_v(m_{v,i})~~~.
\end{equation}
This relation holds in any induced gravity model with vector fields and
does not require additional conditions on the masses of the constituent
fields. It enables one to relate $Q$ to a quantum theory of 2D vector
fields on $\Sigma$.

\section{Discussion}
\setcounter{equation}0

To summarize, we consider a class of induced gravity models
where the low
energy gravitational field is generated by quantum
one-loop effects in a system of heavy constituents. The vector
models presented
here consist of massive scalar, spinor, and vector constituent fields,
and do not require non-minimal couplings of the scalar constituents.
We demonstrated that
the general mechanism of the entropy generation in the induced gravity
proposed in Ref. \cite{FF:97} does work, and the Bekenstein-Hawking
entropy can be derived by statistical-mechanical counting the energy
states of heavy constituents.

It was further demonstrated that the expression for the
Bekenstein-Hawking entropy in the induced gravity can be identically
rewritten in terms of fluctuations of the constituent fields at the
event horizon. The latter are determined only by zero-frequency
(``soft'') solutions of the corresponding field equations. These
``soft'' modes are uniquely defined by their asymptotics at the
bifurcation sphere of horizons $\Sigma$. Using this property, it was
explicitly demonstrated that the fluctuations of the constituent
fields at the horizon coincide with the effective action of
two-dimensional (flucton) fields on $\Sigma$. This mechanism is
somewhat similar to the idea of the holography
\cite{Hooft:93}--\cite{SuWi:98}.
We hope to discuss this relation in more details somewhere
else.

As a result of the two-dimensional reduction, the Bekenstein-Hawking
entropy appears to be equal to $|\lambda|{\cal A}/12$, where $\lambda$
is the 2D cosmological constant induced on the surface of the horizon
by 2D flucton fields. This  implies that the degrees of freedom
responsible for the black hole entropy in the induced gravity can be
related to surface degrees of freedom of the black hole horizon. Such
conclusion is supported by the observation that since the masses of the
constituents are very high (of order of Planckian mass) the
fluctuations of the constituent fields near the horizon can be directly
connected with the fluctuations of the 2D geometry of the horizon.
This might bring connection with the well-known results of statistical
computations of  black hole entropy of 3D black holes \cite{Carlip:95},
\cite{Strominger:97}.

It should be emphasized once again, that the induced gravity approach
does not pretend to explain the black hole entropy from the first
principles of the fundamental theory of quantum gravity (such as the
string theory) but it allows one to demonstrate the universality of the
entropy and its independence of the concrete details of such theory. It
gives us a hint that only a few quite general properties of the
fundamental theory (such as the low energy gravity as induced
phenomenon, finiteness of the low energy coupling constants, holography,
and so on) are really required for the statistical-mechanical
explanation of the black hole entropy.

\vspace{12pt}
{\bf Acknowledgements}:\ \ This work was partially supported  by the
Natural Sciences and Engineering Research Council of Canada. One of the
authors (V.F.) is grateful to the Killam Trust for its financial
support.

\newpage
\appendix
\section{Energy, Hamiltonian, and Noether Charge for Vector Fields}
\setcounter{equation}0

Here we consider the relation between the energy and the Hamiltonian
for vector model and prove Eq. (\ref{2.21a}).
Let us remind that the classical energy $E$ of a field $\Phi$ in a
3D region ${\cal B}$ is
defined by the stress-energy tensor
\begin{equation}\label{2.17a}
E=\int_{\cal B} T_{\mu\nu}\zeta^\mu d\sigma^\nu~~~,~~~
T_{\mu\nu}={2 \over \sqrt{-g}}{\delta I[\Phi] \over \delta g^{\mu\nu}}~~~,
\end{equation}
where $d\sigma^\nu$ is the future directed vector of the volume element on
${\cal B}$, $\zeta^\mu$ is the time-like Killing vector, and
$I[\Phi]$ is the classical action of $\Phi$.
The canonical energy
is
\begin{equation}\label{2.17ab}
H=\int_{\cal B} \left( {\partial L(\Phi) \over \partial \nabla^\nu
\Phi} {\cal L}_\zeta \Phi-
\zeta_\nu  L(\Phi) \right)  d\sigma^\nu~~~,
\end{equation}
where ${\cal L}_\zeta$ is the Lie derivative along $\zeta^\mu$ and
$L(\Phi)$ is the Lagrangian of the field ($I[\Phi]=\int dV L(\Phi)$).

In the quantum theory we are dealing with quantum averages of  the
energy and the canonical energy. For the theory under consideration
(\ref{2.9a}) one has
\begin{equation}\label{ap1}
\bar{E}=\sum_{i=1}^N (\bar{E}_{s,i}+\bar{E}_{d,i}
+\bar{\tilde{E}}_{v,i}-\bar{E}'_{s,i})\, ,
\end{equation}
\begin{equation}\label{ap2}
\bar{H}=\sum_{i=1}^N (\bar{H}_{s,i}+\bar{H}_{d,i}
+\bar{\tilde{H}}_{v,i}-\bar{H}'_{s,i})\, .
\end{equation}
In these relations the average $\bar{C}$ of $\hat{C}$ is
\begin{equation}\label{ap3}
\bar{C}=e^{-i\Gamma[g]}\int [D\Phi] C[\Phi,g] e^{iI[g,\Phi]}\, ,
\end{equation}
and $\bar{E}_i$ and $\bar{H}_i$ are the energy and the canonical energy of each of
the fields which enter total action (\ref{2.10a}). The quantities
$\bar{\tilde{E}}_{v,i}$, $\bar{\tilde{H}}_{v,i}$ and
$\bar{E}'_{s,i}$,
$\bar{H}'_{s,i}$ correspond to the fields
$A^\mu_i$ and $\varphi_i$, respectively. These fields appear under
quantization of the vector constituents $V^\mu_i$.
The minus sign by $\bar{E}'_{s,i}$ and $\bar{H}'_{s,i}$ is the
result of the ''wrong''
statistics of the fields $\varphi_i$\footnote{This result
can be obtained directly
if one starts with the expression for the energy and canonical energy
for a vector field $V_{\mu}$, rewrite them in the point-splitted form,
and takes into account that
$$
\langle\hat{V}_\mu(x)\hat{V}_\nu(x')
\rangle=\langle\hat{A}_\mu(x)\hat{A}_\nu(x')\rangle+
m^{-2}\nabla_\mu\nabla'_\nu\langle\hat{\varphi}(x)\hat{\varphi}(x')
\rangle~~~.
$$}

It can be easily shown that the energy and the canonical energy coincide
if the action $I$ does not contain the terms
with curvature. The only fields which explicitly contain the curvature
term are vector fields $A^\mu_{i}$. For this reason one has
\begin{equation}\label{ap4}
\bar{H}-\bar{E}=\sum_{i=1}^N (
\bar{\tilde{H}}_{v,i}-\bar{\tilde{E}}_{v,i})\, .
\end{equation}
Let us discuss first the difference between the energy $\tilde{E}_v$
and the Hamiltonian  $\tilde{H}_v$
of a classical vector field $A_{\mu}$ described by action $\tilde{I}_v[A]$,
see Eq. (\ref{1.9a}).  $\tilde{E}_v$ and $\tilde{H}_v$ are obtained from
$\tilde{I}_v[A]$ by formulas (\ref{2.17a}) and (\ref{2.17ab}).
The difference between these quantities appears because of the
variation over the metric of the curvature coupling term in $\tilde{I}_v[A]$
so one has
\begin{equation}\label{2.18a}
\tilde{H}_{v}-\tilde{E}_{v}=2\int_{\cal B}d\sigma_\nu
\zeta_\mu\nabla_\rho\nabla_\sigma\left(
{\partial {\tilde L}_v \over \partial R_{\mu\sigma\rho\nu}}
+{\partial {\tilde L}_v \over \partial R_{\nu\sigma\rho\mu}}\right)~~~.
\end{equation}
In the
Rindler approximation the integral can be
transformed to the total
divergence
\begin{equation}\label{2.19a}
\tilde{H}_{v}-\tilde{E}_{v}
=2\int_{\cal B}d\sigma_\nu \nabla_\rho\left[(-\zeta_{\mu;\sigma}
+\zeta_\mu\nabla_\sigma)
\left(
{\partial {\tilde L}_v \over \partial R_{\mu\sigma\rho\nu}}
+{\partial {\tilde L}_v \over \partial
R_{\nu\sigma\rho\mu}}\right)\right]
~~~.
\end{equation}
When the region ${\cal B}$ is the region
of the black hole exterior, the black hole horizon is one of its
boundaries.
Then the integral in r.h.s of (\ref{2.19a}) is reduced to the two terms:
one term comes
from the spatial boundary of $\cal B$ and another one
from the bifurcation surface $\Sigma$.
The terms on the spatial boundary
can be eliminated by the proper choice of the boundary conditions.
However, the term on $\Sigma$ cannot be eliminated.
By taking into account that $\zeta_\mu=0$ and
$\zeta_{\mu;\sigma}=\kappa(t_\mu n_\sigma-
t_\sigma n_\mu)$ on $\Sigma$ one obtains from (\ref{2.19a})
\begin{equation}\label{2.20a}
\tilde{H}_{v}-\tilde{E}_{v}
=4\kappa \int_{\Sigma}t_\mu n_\nu t_\lambda n_\rho
{\partial \tilde{L}_v \over
\partial R_{\mu\nu\lambda\rho}}d\sigma~~~.
\end{equation}
By summing over all the vector fields $A_{i,\mu}$ which enter the model,
using (\ref{ap4}), and
comparing this with Eqs. (\ref{2.11a}) and (\ref{2.12a}) we see
that for the vector induced gravity model
\begin{equation}\label{2.21ab}
H-E={\kappa \over 2\pi} Q~~~,
\end{equation}
where $Q$ is the Noether charge.  To obtain from (\ref{2.21ab}) the
result
in the quantum theory one has to replace the quantities in this formula
by the corresponding quantum averages.

\end{document}